\documentstyle[12pt,psfig]{article}
\newcommand{\be}{\begin{equation}}
\newcommand{\ee}{\end{equation}}
\newcommand{\bea}{\begin{eqnarray}}
\newcommand{\eea}{\end{eqnarray}}
\newcommand{\nnu}{\nonumber}
\newcommand{\norsc}{\normalsize\sc}
\newcommand{\norsl}{\normalsize\sl}
\newcommand{\dd}[1]{\stackrel{\cdot\cdot}{#1}}
\newcommand{\scp}{\scriptstyle}
\newcommand{\ler}{\stackrel{\scp <}{\scp\sim}}
\newcommand{\ger}{\stackrel{\scp >}{\scp\sim}}
\def\p{\phi}
\def\l{\lambda}
\def\L{\Lambda}
\def\e{\epsilon}
\def\d{\delta}
\def\D{\Delta}
\def\pc{\p_{\rm c}}
\def\nm{N_{\rm min}}
\def\x{\xi}
\begin{document}
\begin{titlepage}
\title{\vskip 1.4cm \Large\bf {The No-Boundary Wave Function and the Duration
of the Inflationary Period }}
\author{
\norsc A. Lukas \\
\\
\norsl  Physik Department\\
\norsl  Technische  Universit\"{a}t M\"{u}nchen\\
\norsl  D-85747 Garching, Germany\\ }
\date{}
\maketitle
\begin{abstract}
For the simplest minisuperspace model based on a homogeneous, isotropic
metric and a minimally coupled scalar field we derive analytic expressions
for the caustic which separates Euklidean and Minkowskian region and
its breakdown value $\p_*$. This value represents the prediction of
the no-boundary wave function for the scalar field at the beginning of
inflation. We use our results to search for inflationary models
which can render the no-boundary wave function consistent with the
requirement of a sufficiently long inflationary period.
\end{abstract}
\begin{picture}(5,2)(-300,-400)
\put(2.3,0){TUM--HEP--206/94}
\put(2.3,-20){gr-qc/9409012}
\put(2.3,-40){August 1994}
\end{picture}
\thispagestyle{empty}
\end{titlepage}
\setcounter{page}{1}
Quantum cosmology~\cite{hall} can shade new light on the problem of initial
conditions in the universe. To be predictive boundary conditions have to be
imposed on the wave function of the universe. What quantum cosmology can
do is therefore to replace the procedure of imposing specific boundary
values by a ``theory of boundary conditions''. Several proposals for such
a theory have been made. The most prominent ones are the no-boundary
proposal due to Hartle and Hawking~\cite{hh} and the tunneling proposal
due to Linde and Vilenkin~\cite{lv}. Clearly, all these proposals are
speculative and have to be confronted with reality i.~e.~with
the main characteristics of {\em our} universe.

In the simplest minisuperspace model which is based on a homogeneous,
isotropic metric with scale factor $a$ and a single, minimally coupled
scalar field $\p=\p (t)$ both the no-boundary and the tunneling wave
function predict a correlation between $a$, $\p$ and their conjugate
momenta if $a$ is sufficiently large. Assuming decoherence this correlation
corresponds to a classical universe in a de-Sitter phase. They are therefore in
agreement with the classical behaviour of the universe we live in and the
inflationary paradigm of cosmology. In addition, a probability
distribution $P(\p )$ for the field $\p$ at the beginning of the
inflationary period can be extracted from the wave function. It can be used
to compute a conditional probability $P_{\rm suff}$ for sufficient
inflation~:
\be
 P_{\rm suff} = \frac{\int_{\p_{\rm suff}}^{\p_2}P(\p)d\p}
            {\int_{\p_1}^{\p_2}P(\p)d\p} \; .
 \label{prob}
\ee
Here $\p_{\rm suff}$ is the minimum value of $\p$ which guarantees the
$\nm\simeq 60$ e-folds of inflation needed to solve the problems of
standard cosmology. Since the wave function has been determined in a
semiclassical approximation one should cut off the integral at the value
$\p_2$ corresponding to Planck energy. $\p_1$ is the minimal $\p$
which leads to an expanding universe.

For the tunneling wave function the probability distribution
\be
 P_{\rm T}(\p ) = \exp\left( -\frac{2}{3V(\p )}\right)
\ee
prefers large values of $\p$ and therefore generically predicts
sufficient inflation. However, the opposite is true for the no-boundary
result
\be
 P_{\rm NB}(\p ) = \exp\left(\frac{2}{3V(\p )}\right)\; .
\ee
Since conventionally the potential $V$ contains a tiny coupling in order
to keep the density fluctuations small enough $P_{\rm NB}$ is so
strongly peaked on low values of $\p$ that it essentially predicts the
lower cutoff value $\p_1$ which is much smaller than $\p_{\rm suff}$.
Clearly, this argumentation can be doubted since the behaviour
of $P(\p )$ beyond Planck energy where the above expressions are not valid
may change the picture completely. It has, however, been argued that the
one loop contribution to the distribution $P(\p )$ can suppress the
constant tail at large $\p $~\cite{barv}. This can render $P(\p )$
normalizable and serve as a justification for the semiclassical
approximation. Here we assume that such a suppression indeed
takes place which, in addition, leaves the low energy behaviour of the
probability distribution - as described by the above equations -
essentially unchanged. Furthermore, we assume that we stay in a regime
where the back reaction of quantum fluctuations can be neglected and the
field $\p$ follows a classical slow roll trajectory. Then $P_{\rm suff}$
should indeed be large for a successful inflationary model and its
computation provides a check for a certain boundary proposal.

Grishchuk and Rozhansky~\cite{gr} have pointed out that the existence of
a caustic in the $a$-$\p$-plane is a necessary condition for a
transition from the Euklidean region at small $a$ to the classical
Minkowskian region at large $a$. For small $\p$ this caustic
breaks down so that below a certain value $\p_* >\p_1$  no classical
universe develops. Therefore, in eq.~(\ref{prob}) $\p_1$ should be
replaced by $\p_*$. Then the no-boundary wave function predicts
$\p_*$ as the value of $\p$ at the beginning of inflation. This value
strongly depends on the potential and it has to be clarified whether
$\p_*$ can be larger than $\p_{\rm suff}$ for some choices.

In ref.~\cite{gr} $\p_*$ has been determined numerically for the
potentials $V=\l\p^2$ and $V=\l\p^4$. In both cases it was found to be smaller
than $\p_{\rm suff}$. However, it was left as an open problem whether or not
a potential with $\p_* >\p_{\rm suff}$ exists for the no-boundary wave
function. This is the question we are going to address in this paper.
As a first step we will derive an equation for the caustic and
analytic expressions for $\p_*$. These expressions allow us to discuss the
general dependence of $\p_*$ on the potential $V(\p )$ and to analyze several
classes of potentials in detail.\\

\begin{figure}

\centerline{\psfig{figure=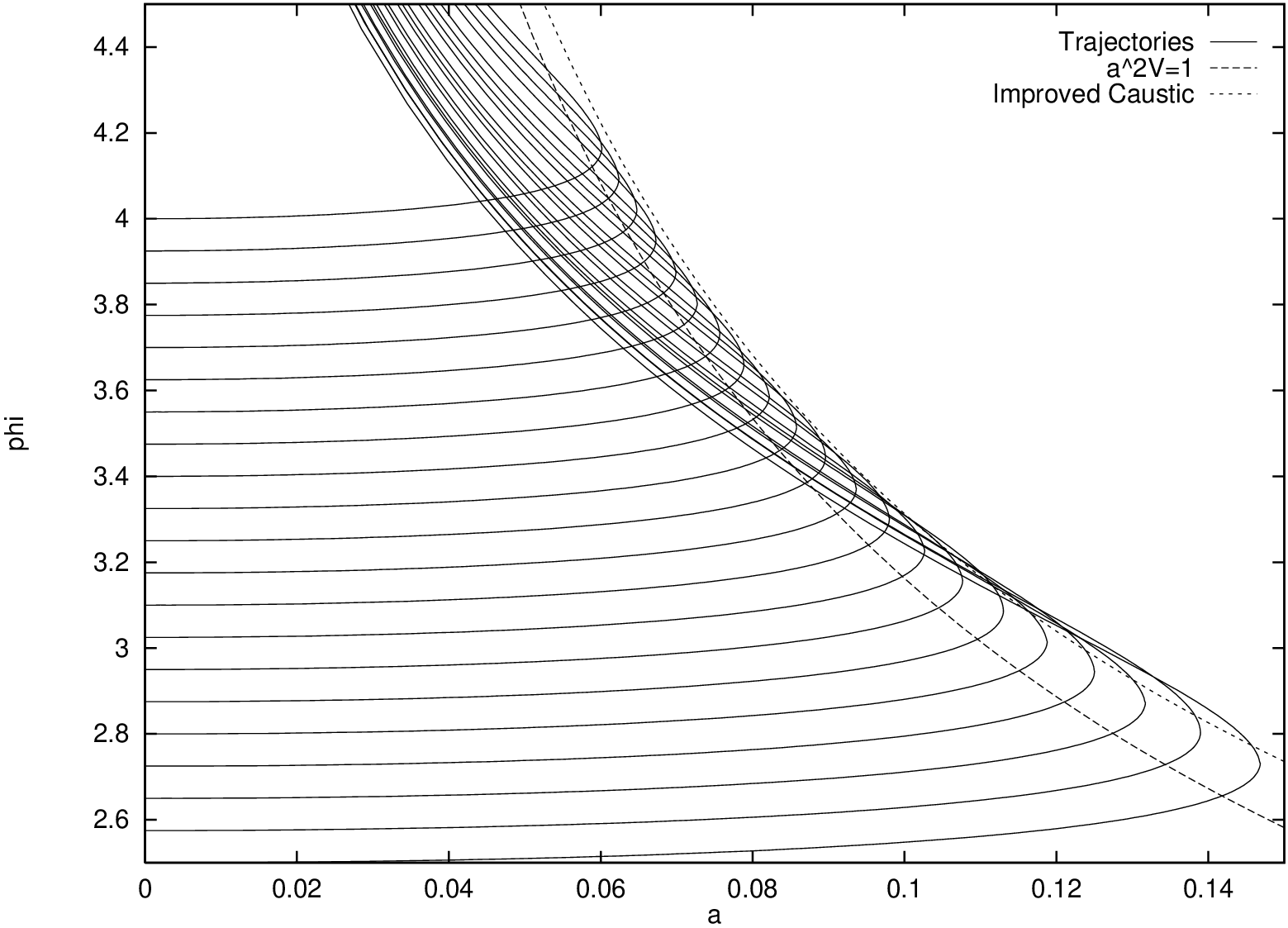,height=12cm,width=10cm,rheight=8.5cm,rwidth=10cm}}
 \caption{Trajectories for the potential $V=\p^4$, approximate caustic
          $a^2V=1$ (dashed line) and improved caustic (dotted line).}
 \label{fig1}
\end{figure}

The Euklidean equations of motion for $a$ and $\p$ which have to be solved
in order to determine the semiclassical wave function are
\bea
 \dd{\p} &+& 3\frac{\dot{a}}{a}\dot{\p} = \frac{1}{2}\frac{dV}{d\p}
  \nnu \\
 \dot{a}^2 &=& 1-a^2V+a^2\dot{\p}^2 \label{eom} \\
 \frac{\dd{a}}{a} &=& -2\dot{\p}^2-V \nnu \; .
\eea
According to the no-boundary proposal they have to be integrated by using
the initial conditions
\[
 a(\tau=0)=0\; ,\quad\quad\dot{\p}(\tau =0)=0\; ,
 \quad\quad\p (\tau =0) =\p_0\; .
\]
To illustrate the qualitative behaviour in fig.~\ref{fig1} we have plotted the
trajectories for a number of starting values $\p_0$ using the
potential~\footnote{The value of $\p_*$ does not
depend on the coupling $\l$ which we have set to $1$ for the numerical
integration.} $V=\l\p^4$. The caustic consists of the points of maximal
$a$ and in this specific example it breaks down at $\p_*\simeq 3$. For large
values of $\p$ it behaves asymptotically like
\be
 a^2V = 1 \label{app_cau}
\ee
(dashed line in fig.~\ref{fig1}) as it can be expected from the second
eq.~(\ref{eom}). Near the breakpoint, however, it deviates substantially from
this asymptotic expression. As a first step we will now improve
eq.~(\ref{app_cau}) such that it also describes the caustic for small $\p$.

Obviously, informations about the time dependency of $a$ and $\p$ are not
needed for this. Instead, we can look at the equation
\be
 (1-a^2V)\p ''+\left[\frac{1}{a}(3-4a^2V)-a(2-3a^2V){\p '}^2\right]\p '
   = \frac{1-a^2{\p '}^2}{2}\;\frac{dV}{d\p} \label{eoma}
\ee
for $\p =\p (a)$ ($\p '=d\p /da$) which can be derived from
eq.~(\ref{eom}). Surprisingly, this equation has two {\em exact} solutions
\be
 \p ' =\pm\frac{1}{a} \label{ex_sol}
\ee
which are independent on the potential $V$. It turns out that the second
of these solutions ($\p ' = -1/a$) asymptotically describes the
``return trajectories'' i.~e.~the trajectories after the
turning point at maximal $a$. Since we are mainly interested in the
behaviour near the approximate caustic $a^2V=1$ we linearize
eq.~(\ref{eoma}) in $\e = a^2V_c-1$, $V_c = V(\pc)$ to obtain
\be
 \e\frac{d\xi}{d\e}\simeq 2\left(\xi^2-\frac{1}{4}\right)
                          \left(\xi +\frac{1}{2y_c}\right)\; ,
 \quad\xi = \frac{d\p}{d\e} \; . \label{leom}
\ee
We have introduced the variable $y$
\be
 y = \frac{2V}{V'}\; ,\quad V' = \frac{dV}{d\p}\; ,\quad
 y_c = y(\pc)\; . \label{y_def}
\ee
which will be quite useful in the following. Eq.~(\ref{leom}) can be
integrated explicitly to give
\bea
 G(\x ) &=& \frac{|\e |}{\e_c (y_c)} \label{integr} \\
 G(\x ) &=& \left( \x +\frac{1}{2y_c}\right)^\frac{2y_c^2}{1-y_c^2}
            \left(\x +\frac{1}{2}\right)^\frac{y_c}{y_c^2-1}
            \left|\x -\frac{1}{2}\right|^\frac{y_c}{y_c^2+1} \label{G_def}
\eea
with
\be
 \e_c (y_c) = y_c^\frac{2y_c^2}{1-y_c^2} \label{eps}
\ee
where we have imposed the boundary condition $\x =0$ at $\e =-1$. At
$\e =\e_c (y_c)$ our solution shows a singularity $\x\rightarrow\infty$
which we identify with the turning point of the trajectory. From this
observation we get for the caustic~:
\be
 a_c^2V_c\simeq 1+\e_c (y_c) \label{cau} \; .
\ee
In fig.~\ref{fig1} this curve is shown for our example $V=\p^4$ (dotted line).

To determine $\p_*$ we have to distinguish two cases. Let us first assume that
the potential in the range we are interested in is increasing rapidly.
This is e.~g.~fulfilled for a polynomial potential like the one in
fig.~\ref{fig1}. Then the first part of the trajectories is approximately
constant in $\p$ and because of the strong variation of the caustic with
$a$ the trajectories follow their asymptotic behaviour $\p ' =-1/a$ even
before they intersect the caustic for the second time.
Then an expression for $\p_*$ can be computed by the following prescription.
A return trajectory defined by $\p ' = -1/a$ which starts at some point
$(a_c,\pc)$ on the caustic~(\ref{cau}) intersects the approximate
caustic $a^2V=1$ at $(a,\p )$. Using the eqs.~(\ref{app_cau}),
(\ref{ex_sol}), (\ref{cau}) and (\ref{eps}) we find the relation
\be
 \frac{e^\p}{\sqrt{V(\p)}} = \sqrt{1+\e_c (y_c)}\;\frac{e^{\pc}}
                             {\sqrt{V(\pc)}}\; .
\ee
Now we minimize $\p$ as a function of $\pc$ to identify the lowest
return trajectory (cf.~fig.~1). The value of $\p (\pc )$ at the minimum
can be identified with $\p_*$. The explicit computation
results in conditions on $y_c=y(\pc )$ and $\p_*$~:
\bea
 \frac{1}{y_c}&-&\frac{1}{2}\;\frac{d\e_c /dy}{1+\e_c}(y_c)\;
   y' (y_c) = 1 \label{c1} \\
 \frac{e^{\p_*}}{\sqrt{V(\p_*)}} &=& \sqrt{1+\e_c (y_c)}\;
 \frac{e^{\pc}}{\sqrt{V(\pc)}}\; . \label{c2}
\eea
{}From eq.~(\ref{eps}) it can be seen that $\e_c (y=1) = e^{-1}$ and
$d\e_c /dy (y=1) = -e^{-1}$. Therefore, if
\be
 y' (y) = 2\left(1-\frac{1}{4}y^2\frac{V''}{V}(y)\right) \label{yp}
\ee
is $\ler O(1)$ at $y=1$ we can consistently neglect the second term in
eq.~(\ref{c1}). We will see that this is indeed justified for a number of
potentials. The above equations then simplify to
\bea
 y_c &=& \frac{2V}{V'}(\pc) = 1 \label{ca1} \\
 \frac{e^{\p_*}}{\sqrt{V(\p_*)}} &=& k\;
 \frac{e^{\pc}}{\sqrt{V(\pc)}}\; . \label{ca2}
\eea
with
\[
 k = \sqrt{1+e^{-1}}\simeq 1.17 \; .
\]
These equations now allow an explicit calculation of $\p_*$ for a given
sufficiently increasing potential. Since the constant $k$ in eq.~(\ref{ca2})
is close to unity $\pc$ and $\p_*$ are of the same order for almost all
potentials~\footnote{An exception to this is given by an exponential
potential which cancels the exponentials in the numerators of eq.~(\ref{ca2})
to a high accuracy. We have discussed this special case in an example below.}.
An order of magnitude estimate for $\p_*$ is therefore given by the solution
to the simple equation
\be
 y = \frac{2V}{V'} = 1 \; .
\ee
{}\\

\begin{figure}

\centerline{\psfig{figure=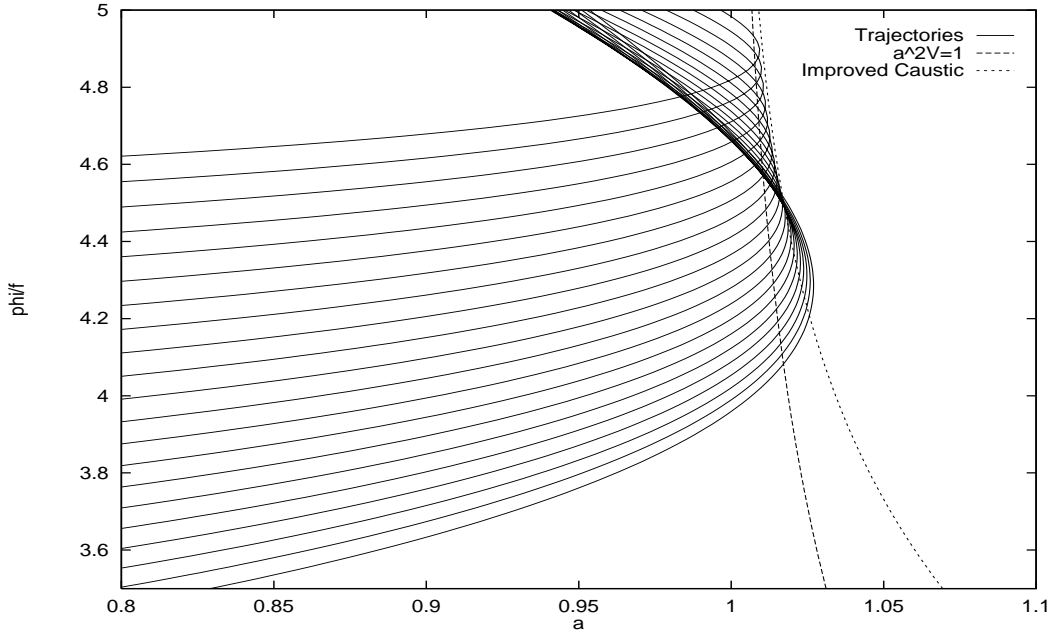,height=12cm,width=10cm,rheight=8.5cm,rwidth=10cm}}
 \caption{Trajectories for the potential $V=(1-\exp(-\p /f))^2$ for $f=0.1$,
          approximate caustic  $a^2V=1$ (dashed line) and improved caustic
          (dotted line).}
 \label{fig2}
\end{figure}

For potentials with a very flat region e.~g.~caused by a maximum or an
asymptotic behaviour $V\rightarrow$ const the situation is different.
Fig.~\ref{fig2} shows the example $V=(1-\exp(-\p /f))^2$ for $f=0.1$.
The trajectories increase substantially from $a=0$ up to the caustic which
is nearly vertical. The return up to the second intersection with the caustic
is too short to be described by the asymptotic equation $\p '=-1/a$.
Instead, we follow a trajectory starting out at $y_0=y(\p_0 )$, $a=0$ and
intersecting $a^2V=1$ at $y_c=y(\pc)$ by using the equations
\bea
 \x &=& \frac{1}{6y_0}\quad\quad a\; {\rm small} \nnu \\
 \x &=& \frac{1}{2}\sqrt{\frac{\e_c(y_c)}{\e_c(y_c)-\e}}\quad\quad
        \e\; {\rm small} \nnu \; .
\eea
They can be derived from eq.~(\ref{eoma}) and from eq.~(\ref{integr}) by
expanding $G(\x )$ up to the first order in $1/\x$. These approximate
first integrals for small $a$ and the region around the caustic can be
``glued'' together. Their integration gives the second intersection value
$\p$ with the caustic $a^2V=1$ as a function of $\p_0$. As before the minimum
of this function should be identified with $\p_*$. Following this
prescription we arrive at a set of equations
\bea
 \p_* &=& \pc +\frac{2}{y_c^2} \nnu \\
 \pc &=& \p_0 +\frac{1}{3y_0}-\frac{y_0}{y'(y_0)} \label{ca_flat} \\
 \frac{3}{2y_c^2} &=&\frac{1}{6y_0^2}-\frac{1}{y'(y_0)} \nnu
\eea
where we have used that $\e_c (y) \simeq 1/y^2$ since $y\gg 1$ in a flat
region of the potential.\\

For the two types of potentials - the increasing ``chaotic'' type of potentials
and the type with a flat region - the eqs.~(\ref{c1}), (\ref{c2}) and
eq.~(\ref{ca_flat}) represent the general answer for the value of $\p_*$.
Clearly, we have not {\em proven} that our approximations
are justified for all potentials. The results should therefore be used as
a guide to find promising potentials whose properties then have to be
verified numerically. In this way large classes of potentials can be
analyzed without a numerical case by case study. In fact, for all potentials
we have checked so far our analytic formulae gave the correct answer.\\

We start by discussing the consequences of the previous results on the
first type of potentials. The number of e-folds in the slow roll
approximation is given by
\be
 N_{\rm e-folds} = 3\int_{\p_*}^{\p_f}y(\p )\, d\p \label{e_folds}
\ee
where $\p_f$ is the end point of the slow roll regime determined by the
breakdown of the conditions
\be
 y\gg \frac{1}{3}\; ,\quad\frac{V''}{V}\ll 18 \; . \label{end_infl}
\ee
Suppose we have found a potential $V$ of the first type which generates the
desired number of e-folds. Since $y$ does not increase ``irregularly'' it is
plausible that the slow roll conditions~(\ref{end_infl}) are still fulfilled
at $\pc$. Consequently, our approximation leading to~(\ref{ca1})
and~(\ref{ca2}) is justified and we have $y_c\simeq 1$ at $\pc$.
This leads to the rough approximation
\be
 N_{\rm e-folds} \ler 3\p_* \; .
\ee
It shows that for such potentials one will need a quite large value $\p_*$
to generate the needed $\nm\simeq 60$ e-folds. From $y(\p )<1$ for
$\p\ler \p_*$ we conclude that
\[
 \frac{V(\p_2)}{V(\p_1)} > e^{2(\p_2 -\p_1 )}
\]
for any $\p_2$, $\p_1 \ler\p_*$. In particular we have
\be
 V(\p_* )\ger\l e^{2\p_*}\sim\l e^\frac{2N_{\rm min}}{3}\sim\l e^{40}
\ee
where $\l$ is the coupling in front of $V$. Therefore a potential which
can provide a large enough $\p_*$ either exceeds Planck energy at this
point - which makes the predicted $\p_*$ unreliable - or needs a very small
coupling $\l$. In the second case quantum fluctuations in de Sitter space
are generically too small to generate a sufficiently large initial
density spectrum. Then other sources of such density perturbations have
to be considered.

Examples are provided by the standard chaotic potentials
$V=\l\p^n$, $n=2,4,6,\cdots$. It is easy to show that $y' (y=1)=2/n$ so
that the eqs.~(\ref{ca1}) and~(\ref{ca2}) can be applied. We get
\be
 \pc = \frac{n}{2} \; .
\ee
Writing $\p_* = \pc +\d$ for sufficiently large $n$ we can approximate
$\sqrt{V(\p_*)}\simeq\sqrt{V(\pc )}(e(1-\d /n))^\d$ to obtain an explicit
solution of eq.~(\ref{ca2})~:
\be
 \p_* \simeq \frac{n}{2}+\sqrt{1-k^{-1}}\,\sqrt{n}
      \simeq \frac{n}{2}+0.38\sqrt{n} \; .
\ee
By numerical integration of eq.~(\ref{eom}) we have checked up to $n=70$
that this formula indeed results in reasonable values for $\p_*$. The
number of e-folds is given by
\[
 N_{\rm e-folds}\simeq\frac{3}{n}\,\p_*^2\simeq\frac{3}{4}\,n+1.14\sqrt{n} \; .
\]
In accordance with the general discussion a very large value of $\p_*$ and
correspondingly of $n$ is needed to generate sufficient inflation.
In fact this is possible for $V=\l\p^n$ with $n\ger 70$ but an enormously
small $\l$ is needed to keep $V(\p_* )$ below Planck energy.

Let us next discuss the potential
\be
 V = \L^4 \left(1-\exp\left(\frac{\p}{f}\right)\right)^2 \label{hog}
\ee
which (for $f=1$) can be derived from higher order gravity theories~\cite{hog}.
We concentrate on the interesting range $f\ger 1$ where the exponentials
in eq.~(\ref{ca2}) are canceled. In that case $\p_*$ may be substantially
larger than $\pc$ and the above general argument is somewhat weak.
{}From the eqs.~(\ref{ca1}), (\ref{ca2}) and (\ref{e_folds}) we obtain for
$0<d =f-1\ll 1$~:
\bea
 \pc &\simeq& \ln\frac{1}{d} \nnu \\
 \p_* &\simeq&\ln\frac{1}{d} + \frac{\ln k}{d} \\
 N(\p_* ) &\simeq& 3\p_* \nnu \; .
\eea
These formulae have been confirmed by numerical integration. For $d\ler 0.01$
a sufficient number of e-folds may be generated, however, in accordance with
our general conclusion a small coupling $\L^4\simeq e^{-40}$ is needed to
keep $V(\p_* )$ below the Planck scale.\\

Now we turn to the second type of potentials with a flat region.
An interesting example is provided by the above potential~(\ref{hog})
in the negative $\p$ direction where it approximates a constant value.
Quantum cosmology of this potential has been studied in a number of
papers including~\cite{hog_qc}. Only for a value of $\p_*$ in the
asymptotically flat region we can hope for sufficient inflation. Such values
are obtained for $f\ll 1$ where we can solve the system~(\ref{ca_flat})
explicitly. We find~:
\be
 \frac{\p_*}{f}\simeq\ln\frac{1}{1.45f^2}+4.20 f\; .
\ee
This has been confirmed numerically. Given the formula
\[
 N_{\rm e-folds}(\p_* ) \simeq 3f^2\exp\left(\frac{\p_*}{f}\right)
\]
we see that independent on $f$ only a few e-folds are generated starting
at $\p_*$.

Another way to obtain a flat region is to consider a potential with
a maximum at some value $\p =\p_m$. A sufficiently long period of
inflation will be produced for these potentials if $\p$ starts its
evolution in a small range around the maximum. Basically, we have to
ask two kinds of questions~: Can the shape of the potential be such
that $\p_*$ falls into this range and is there a fine tuning involved in
arranging such a situation ?

Before we address these questions in general let us be more specific
and discuss a prominent example for these potentials~:
The natural inflation potential $V=\Lambda^4 (1-\cos (2\p /f))$ which
can be interpreted as the effective potential of a pseudo-goldstone boson.
Its inflationary properties have been studied extensively in
ref.~\cite{nat}. The quadratic maximum of this potential is still ``curved''
enough to apply the eqs.~(\ref{ca1}) and~(\ref{ca2}). From
\be
 \cos\left(\frac{2\pc}{f}\right) = \frac{f^2-1}{f^2+1}
\ee
we see that $\pc$ and correspondingly $\p_*$ approach the maximum
for decreasing $f$. The latter is found from a numerical solution of
eq.~(\ref{ca2}) and is in agreement with the results from numerical integration
of~(\ref{eom}). At $f\simeq 0.6$ the value $\p_*$ reaches the maximum.
In principle, one can therefore hope that a value of $f$ close to $0.6$\
produces sufficient inflation. However, taking into account that
\be
 \pi -\frac{2\p_{\rm suff}}{f} \sim \exp\left(-\frac{\nm}{3f^2}\right)
\ee
this seems almost impossible~: For $f\simeq 0.6$ the starting point of
the field $\p$ has to be at the maximum to such an enormous precision that
one has to fine tune $f$ in order to obtain a $\p_*$ in this range.
Even if such a tuning is accepted one would have to worry about the
fluctuations of the higher, inhomogeneous modes and their back reaction
on the field $\p$. This argument does not depend on the specific choice
of the potential but just on the quadratic behaviour around the maximum.
Our conclusion holds for all potentials with such a quadratic maximum
provided that the slow roll near the maximum is the main source
of inflation.

We can generalize this analysis to higher order maxima.
If we assume that the relevant phenomena appear near the maximum we can
expand the potential around $\p_m$ up to the first nontrivial order $2n$~:
\be
 V = \L^4 (1-\D^{2n})+O(\D^{2n+1})\; ,\quad\D = \frac{\p_m -\p}{f}\; .
 \label{max_pot}
\ee
For the number of e-folds we get
\be
 N_{\rm e-folds} (\D_*) \sim \frac{3f^2}{2n(n-1)}\frac{1}{\D_*^{2n-2}}
 \quad\quad{\rm for}\; n>1 \; .
\ee
In contrast to the $n=1$ case the number of e-folds does not show a
logarithmic dependence on the initial value so that fine tuning
problems are irrelevant. The equations~(\ref{ca_flat}) can be solved
with the ansatz $y_0 = d_n/f^{n/(n-1)}$, $y_c = c_n/f^{n/(n-1)}$ leading to
\bea
 \D_* &\simeq& (nc_n)^{-\frac{1}{2n-1}}f^\frac{1}{n-1}  \\
 N_{\rm e-folds}(\D_* ) &\simeq&\frac{3}{2n(n-1)}(nc_n)^\frac{2n-2}{2n-1} \; .
\eea
As for the exponential potential the number of e-folds turns out to be
independent on $f$. The numerical constants $c_n$ are small,
e.~g.~$c_2\simeq 2.6$, $c_3\simeq 5.0$, showing that $N_{\rm e-folds}(\D_* )$
remains below the needed value. For $n=2,3$ this result has been verified
by numerical integration.\\

In conclusion, we have determined an analytic equation for the caustic
separating the Euklidean and Minkowskian region in the simplest
minisuperspace model. Moreover, we derived formulae for the breakdown
value $\p_*$ of the caustic which is the value
of the scalar field $\p$ predicted by the no-boundary wave function.
Our results give some insight in the structure of minisuperspace models
and its dependency on the scalar field potential. They allow to analyze
inflationary properties of these models without relying on a numerical
case by case study. We have applied these results to a wide range of
models with a single scalar field. Unfortunately, for potentials with
an extraordinary flat region, e.~g.~caused by a maximum or an asymptotic
behaviour $V\rightarrow$ const we could not find an example which leads
to a sufficiently long period of inflation. For potentials with a maximum
and an inflationary phase which mostly arises due to the slow roll near the
maximum this behaviour seems to be general.

On the other hand, some potentials of ``chaotic'' type
like $V=\l\p^n$ with $n\ger 70$ can cause sufficient inflation.
However, to stay within the limits of the semiclassical approximation the
coupling $\l$ generically has to be very small. A positive answer
within these limits therefore asks for an alternative explanation
of primordial density perturbations. From our results, one might also argue
that an understanding beyond the semiclassical limit is necessary
to decide about the existence of a ``successful'' model. On the other
hand, we have concentrated on the simplest class of models with a
single scalar field which in the inflationary phase behaves essentially
classical. Non negligible quantum fluctuations, in particular in the
context of multidimensional models, might, however, change the picture.
{}From the viewpoint of particle physics models with more than one
scalar field are the more realistic anyway. Generalizations of the results
presented in this paper might be helpful in analyzing the quantum
cosmology of such models.\\[0.5cm]

{\bf Acknowledgment}~I would like to thank H.~van Elst, Z.~Lalak and R.~Poppe
for helpful discussions. This work was partially supported by the Deutsche
Forschungsgemeinschaft and the EC under contract no.~SC1-CT92-0789 and
the CEC Science Program no.~SC1-CT91-0729.
\end{document}